# Decoupling of magnetism and electric transport in single-crystal $(Sr_{1-x}A_x)_2IrO_4$ (A = Ca or Ba)

H. D. Zhao[1], J. Terzic[1,2], H. Zheng[1], Y. F. Ni[1], Y. Zhang[1], Feng Ye[3] and P. Schlottmann[4] and G. Cao[1*]

[1] Department of Physics, University of Colorado at Boulder, CO 80309

[2] National High Magnetic Field Laboratory, Tallahassee, FL 32306

[3] Oak Ridge National Laboratory, Oak Ridge, TN 37830

[4] Department of Physics, Florida State University, Tallahassee, FL 32306

## Abstract

We report a systematical structural, transport and magnetic study of Ca or Ba doped $Sr_2IrO_4$ single crystals. Isoelectronically substituting $Ca^{2+}$ (up to 15%) or $Ba^{2+}$ (up to 4%) ion for the $Sr^{2+}$ ion provides no additional charge carriers but effectively changes the lattice parameters in $Sr_2IrO_4$. In particular, 15% Ca doping considerably reduces the *c*-axis and the unit cell by nearly 0.45% and 1.00 %, respectively. These significant, anisotropic compressions in the lattice parameters conspicuously cause no change in the Néel temperature which remains at 240 K, but drastically reduces the electrical resistivity by up to five orders of magnitude or even precipitates a sharp insulator-to-metal transition at lower temperatures, i.e. the vanishing insulating state accompanies an unchanged Néel temperature in $(Sr_{1-x}A_x)_2IrO_4$. This observation brings to light an intriguing difference between chemical pressure and applied pressure, the latter of which does suppress the long-range magnetic order in $Sr_2IrO_4$. This difference reveals the importance of the Ir1-O2-Ir1 bond angle and homogenous volume compression in determining the magnetic ground state. All results, along with a comparison drawn with results of Tb and La doped $Sr_2IrO_4$, underscore that the magnetic transition plays a nonessential role in the formation of the charge gap in the spin-orbit-tuned iridate.

[*] Email: gang.cao@colorado.edu

I. **Introduction**

$Sr_2IrO_4$ is the archetype $J_{eff}=1/2$ Mott insulator[1] with a Néel temperature $T_N$ = 240 K [2,3,4,5] and an energy gap $\Delta \leq 0.62$ eV [6,7,8] and a relatively small magnetic coupling energy of 60-100 meV. [7,9] The distinct energy hierarchy featuring a strong spin-orbit interaction (SOI) and the novel $J_{eff}=1/2$ Mott state have motivated a large number of experimental and theoretical investigations on $Sr_2IrO_4$ in recent years.[10]

It is commonly thought that $Sr_2IrO_4$ crystallizes in a tetragonal structure with space-group $I4_1/acd$ (No. 142) with $a = b = 5.4846$ Å and $c = 25.804$ Å at 13 K [2,4] (**Fig.1a**). More recently, studies of neutron diffraction and second-harmonic generation of single-crystal $Sr_2IrO_4$ [11,12] revealed structural distortions and forbidden reflections such as (1, 0, 2n+1) for the space group $I4_1/acd$ over a wide temperature interval, 4 K < T < 600 K, which lead to a further reduced structural symmetry with a space group $I4_1/a$ (No. 88). Nevertheless, an important structural feature of $Sr_2IrO_4$ is a rotation of the $IrO_6$-octahedra about the *c*-axis by ~12° (**Fig.1a**). Furthermore, neutron diffraction reveals a long-range antiferromagnetic (AFM) state below 240 K with an ordered moment of 0.208(3) $\mu_B$/Ir and a canted AFM configuration within the basal plane.[11] Remarkably, these moments deviate 13(1)° away from the *a*-axis, indicating that the magnetic moment canting rigidly tracks the staggered rotation of the $IrO_6$ octahedra.

The present study is motivated by the unconventional correlation between the AFM and $J_{eff}=1/2$ Mott states in $Sr_2IrO_4$ that is manifest in the following empirical trends. **(1)** $Sr_2IrO_4$ exhibits no discernable anomaly in the electrical resistivity at the AFM transition at $T_N$ (= 240 K), indicating an unusual association of the AFM state with the insulating gap[5,10]. **(2)** $Sr_2IrO_4$, along with other iridates, does not metallize at pressure up to 50 GPa [13,14,15,16,17,18] but the long-range AFM order disappears near 20 GPa. The avoidance of a metallic state at high pressures is inconsistent with a common occurrence that an insulating state collapses and a metallic state emerges at high pressures as the unit cell shrinks and the bandwidth broadens. **(3)** In sharp contrast, a growing body of experimental evidence has shown that a metallic state can be readily realized via slight chemical doping, either electron (e.g. oxygen depletion[19], La doping [20,21]) or hole doping (e.g., K[20] or Rh[22] or Ru[23,24] doping, for either Sr or Ir) despite the sizable energy gap (~ 0.62 eV[8]). However, the emerging metallic state does not necessarily accompany a disappearance of the AFM state.[10]

The apparently unusual correlation between the insulating and magnetic properties has helped revitalize discussions of Mott, Mott-Hubbard and Slater insulators, in particular, the dependence of the charge gap formation on magnetic interactions in $Sr_2IrO_4$. [10, 25, 26,27] Indeed, the conspicuous absence of bulk superconductivity, which was widely predicted in $Sr_2IrO_4$, [28] is another testament to such an unconventional correlation despite structural, electronic and magnetic similarities between $Sr_2IrO_4$ and $La_2CuO_4$ (i.e., $K_2NiF_4$ type, one hole per Ir or Cu ion, pseudospin- or spin-1/2 AFM, etc.). Clearly, a better understanding of the $J_{eff}$=1/2 state and its correlation with the AFM state in $Sr_2IrO_4$ needs to be established.

Early studies of doped $Sr_2IrO_4$ often involve simultaneous charge (electron or hole) doping and lattice alterations, such as La, K, Mn, Ru, Rh, Tb, etc. substitutions. While these dopants are proven effective in tuning the ground state, the dependence of magnetic and transport properties on the lattice degrees of freedom cannot be examined without ambiguity. Studies of Ca doped $Sr_2IrO_4$ in single-crystal form[29] and epitaxial thin-film form [30] were also previously reported; results of these studies provide useful reference.

Here, we report a systematical structural, transport and magnetic study of Ca or Ba doped $Sr_2IrO_4$ or $(Sr_{1-x}A_x)_2IrO_4$ single crystals (where A=Ca or Ba). Isoelectronically substituting $Ca^{2+}$ (up to 15%) or $Ba^{2+}$ (up to 4%) ion for the $Sr^{2+}$ ion provides no additional charge carriers but effectively tunes the lattice parameters of $Sr_2IrO_4$. The ionic radii of the three cations are considerably different, namely, 1.00 Å, 1.18 Å and 1.35 Å for Ca, Sr and Ba, respectively, thus substituting Ca (Ba) for Sr is expected to reduce (expand) the lattice. Therefore, any magnetic and electronic responses to Ca or Ba doping must be exclusively due to changes in the lattice parameters. This work reveals that Ca doping considerably compresses the *c*-axis and the unit cell volume by nearly 0.45% and 1.00%, respectively. The significant, anisotropic compressions in the lattice parameters conspicuously cause no change in the Néel temperature at 240 K and the Curie-Weiss temperature, but a drastic reduction in the electrical resistivity by up to five orders of magnitude or even precipitates a sharp insulator-to-metal transition at lower temperatures -- The vanishing insulating state accompanies an unchanged Néel temperature. It suggests that the electron hopping is strongly enhanced as a result of the volume shrinking but the exchange coupling that dictates long-range magnetic order does not seem to be sensitive to the compression in the lattice of $(Sr_{1-x}A_x)_2IrO_4$ although the magnetic anisotropy does change notably. This observation brings to light an interesting difference between chemical pressure and applied pressure, the latter of which does suppress the long-range AFM order at 20 GPa but retains the insulating state in $Sr_2IrO_4$.[13] This difference signals the importance of the in-plane Ir1-O2-Ir1 bond angle in determining the magnetic ground state in $Sr_2IrO_4$. The changes in the Ir1-O2-Ir1 bond angle are noticeably small and the effect of them may be local and non-uniform in $(Sr_{1-x}Ca_x)_2IrO_4$. The results of this work along with the comparison drawn with results of high pressure studies and studies of Tb and La doped $Sr_2IrO_4$ underline a unique, unconventional correlation between the magnetic and $J_{eff}=1/2$ states, which characterizes this archetype $J_{eff}=1/2$ Mott insulator. It is worth mentioning that some results of this work are consistent with those reported in the earlier study on Ca doped $Sr_2IrO_4$[29], but this work *reveals and emphasizes*, among a few other new, distinct features, a unchanged Néel temperature in $(Sr_{1-x}A_x)_2IrO_4$ for both Ca and Ba doping, and an evolution of magnetic and transport anisotropies with Ca or Ba doping in $Sr_2IrO_4$, and a comparison between effects of various dopants and between effects of chemical pressure and applied external pressure in $Sr_2IrO_4$, which are not reported in Ref.29.

## II. Experimental

Single crystals of Ba or Ca doped $Sr_2IrO_4$ were grown using a self-flux method from off-stoichiometric quantities of $IrO_2$, $SrCO_3$, $SrCl_2$ and $BaCO_3$ or $CaCO_3$. The average crystal size of doped $Sr_2IrO_4$ is of 1 x 0.4 x 0.2 mm$^3$ (see **Fig.1b**). The highest doping level is 15 % for Ca but only 4% for Ba. This is not unexpected because $Ba_2IrO_4$ can only form at high pressures. Measurements of crystal structures were performed using a Bruker D8 Quest ECO single-crystal diffractometer that features a PHOTON 50 CMOS detector. It is also equipped with an Oxford Cryosystem that creates sample temperature environments ranging from 100 K to 380 K during x-ray diffraction measurements. Chemical compositions of the single crystals were determined using a set of Hitachi/Oxford MT 3300 Scanning Electron Microscope (SEM) and Energy Dispersive X-ray (EDX). Standard four-lead measurements of the electrical resistivity were carried out using a Quantum Design Dynacool PPMS System equipped with a 14-Tesla magnet. Magnetic properties were measured using a Quantum Design MPMS-7 SQUID Magnetometer.

## III. Results and Discussion

Results of our crystal structural study of single-crystal $(Sr_{1-x}Ca_x)_2IrO_4$ ($0 \leq x \leq 0.15$) and $(Sr_{1-x}Ba_x)_2IrO_4$ ($0 \leq x \leq 0.04$) are summarized in **Fig.2**. The lattice parameters systematically decrease with Ca doping x, as anticipated (**Figs.2a, 2b** and **2c**). The *a*- and *c*-axis are shortened by up to 0.22% and 0.45 %, respectively, whereas the unit cell volume V shrinks by an astonishing 0.9% for x (Ca) = 0.12 at 100 K (**Fig.2d**). It is emphasized that Ca doping causes a more significant compression in the *c*-axis than in the *a*-axis, which helps enhance the inter-plane coupling and modify magnetic and electronic anisotropies, as discussed below. On the other hand, the relatively low Ba doping level introduces little changes in the lattice parameters *a*-, *c*-axis and V in general. The Ir1-O2-Ir1 bond angle in the basal plane changes only slightly and cannot be precisely determined because of the large error bar (in contrast, the lattice parameters *a*-, *c*-axis and V can be nearly precisely determined; the related error bars are negligible.); the data in **Fig.2e** seem to suggest a more noticeable change in the $IrO_6$ rotation with Ba doping than with Ca doping.

It is surprising that the Néel temperature $T_N$ = 240 K remains essentially unchanged for both Ca and Ba doped $Sr_2IrO_4$, as shown in **Figs.3a** and **3b**, despite the significant changes in the lattice parameters. Furthermore, the Curie-Weiss temperature, $\theta_{CW}$, and effective moment $\mu_{eff}$, both of which are extrapolated from the inverse magnetic susceptibility $\Delta\chi^{-1}$ ($\Delta\chi = \chi(T) - \chi_o$, where $\chi_o$ is a T-independent contribution, also show little changes throughout the entire Ca and Ba doping range (**Fig.3a** right scale, and **Fig.3c**). Note that given the presence of the AFM in $Sr_2IrO_4$, it is intriguing that $\theta_{CW}$ = +236 K is positive for x = 0, [5] and remains unchanged as well in Ba or Ca doped $Sr_2IrO_4$. Conventionally, a magnetically ordered state is closely related to the Heisenberg exchange parameter *J*, which is given approximately by $J = -2t^2/U$, where *U* is the on-site Coulomb interaction and the hopping integral *t* is proportional to the band width. The unchanged $T_N$ seems to suggest an unaffected *t* in the Ca and Ba doped $Sr_2IrO_4$. Indeed, the AFM correlation length $\xi$ is roughly given by $k_F\xi \sim E_F/T_N$ (where $k_F$ and $E_F$ are Fermi wavevector and Fermi energy, respectively). An estimated $\xi$ is approximately 17-20 lattice spacing long; a sphere of such a coherence length ξ contains about 700 unit cells; if 10% of them have scattering centers due to Ca doping, the effect is inconsequential -- the AFM state is stable. However, $T_M$ near 100 K, a signature feature for pure $Sr_2IrO_4$, disappears upon Ca doping or weakens due to 2% of Ba doping (**Fig.3a**). $T_M$ is attributed to a moment reorientation or magnetic canting that is at the root of the unusual magnetoresistivity[20] and giant magnetoelectric behavior[31] observed in $Sr_2IrO_4$.[32] The vanishing $T_M$ implies that the changes, however small, in the Ir1-O2-Ir1 bond angle effectively modify the magnetic canting. The above results are noticeably different from those reported in the Ref. 29, where $T_N$ changes with Ca doping (no data for Ba doping in Ref. 29).

Indeed, Ba and Ca doping does alter the magnetic anisotropy more significantly. This is manifest in **Figs. 3a**, **3b** and **Fig.4**. The magnetic anisotropy between the *a*- and *c*-axis is reduced with Ca doping in part because of the shorten c-axis, which enhances the *c*-axis magnetization (**Figs.4a** and **4b**). For example, the ratio of $M_a/M_c$ at 7 T and 1.8 K reduces from 2.5 for x = 0 to 1.2 for x = 0.12 (see **Fig.4c**). The reduced magnetic anisotropy can be attributed to the significantly compressed *c*-axis (by ~ 0.45%, **Fig.2d**), which enhances the interlayer-coupling. Ba doping seems to retain or slightly enhance the magnetic anisotropy although the Ba doping level is low **(Figs. 3a**, **3b** and **Fig.4**).

The absence of any changes in the long-range AFM order in Ca or Ba doped $Sr_2IrO_4$ is striking, particularly when it is compared to the effect of applied pressure on $Sr_2IrO_4$. Chemical doping that causes expansion or compression of a unit cell is often referred to as *chemical pressure*. The chemical pressure due to Ca doping considerably compresses the unit cell volume V (by ~1%) but causes no discernible changes in $T_N$. However, application of pressure does suppress the long-range AFM order near 20 GPa.[13] This stark difference between chemical pressure and applied pressure in $Sr_2IrO_4$ is schematically illustrated in **Fig.4d**. The central effect of applied pressure on all solids is a shrinking of interatomic distances and volume as a result of bond compression and bond-angle bending. In terms of energy scale for the present case, 10 GPa is approximately equal to 0.1 eV/Å$^3$. The bond energy of iridates is high and of a few electron volts (high melting temperatures (>1900 °C) of these materials are a result of the strong bond energy). A gross estimate suggests that 1.0% of volume compression in the iridate might require approximately 5 GPa assuming ~ 0.1% bond compression/GPa, given the relatively incompressible nature of the iridate (for comparison, this value ~ 0.07% for diamond and < 0.2% for silicates). Application of pressure of 5 GPa is not sufficient to completely destabilize the AFM state, according to Ref. 13, but it is strong enough to suppress $T_N$ in a sizable way. The fact that no change in $T_N$ occurs in response to the compressed volume in Ca doped $Sr_2IrO_4$ might be due to the following two reasons, namely, anisotropic compression, and insignificant and local changes in the Ir1-O2-Ir1 bond angle. The compression rate of the *c*-axis is almost twice as much as that of the *a*-axis in Ca doped $Sr_2IrO_4$ (**Fig.2d**), leading to a reduced ratio of the *c*-axis to the *a*-axis, *c/a* (=4.709 for x=0 and 4.698 for x=0.15 at 100 K), which may help enhance the inter-plane magnetic coupling. This may be consistent with the higher $T_N$ = 285 K in the bi-layered $Sr_3Ir_2O_7$ [33]. The difference between chemical and applied pressure could also originate from the changes in the in-plane Ir1-O2-Ir1 bond angle: chemical pressure due to Ca doping causes slight or little changes which are likely non-uniform and local, whereas applied pressure can introduce significant changes in the Ir1-O2-Ir1 bond angle or bond-angle bending that are uniform or global in the sample.

In contrast to the magnetic response to Ca or Ba doping, the transport properties change drastically with Ca or Ba doping. The electrical resistivity $\rho$ decreases systematically with increasing Ca or Ba doping, as shown in **Fig. 5**. The ratio of *c*-axis $\rho_c$ to *a*-axis $\rho_a$, $\rho_c/\rho_a$, is reduced from $10^3$ for x=0 to less than 10 for Ca doping at x =0.12 but remains significantly large for Ba doped $Sr_2IrO_4$. More significantly, both $\rho_a$ and $\rho_c$ reduce by as much as several orders of magnitude at low temperatures upon Ca or Ba doping. But the decreases in $\rho_a$ and $\rho_c$ are noticeably different in magnitude. For example, the magnitude of low temperature $\rho_c$ reduces by five orders of magnitude, from $10^6$ Ω cm for x = 0 to 10 Ω cm for x = 0.12 for Ca doping, whereas this value is only three orders of magnitude for $\rho_a$. The larger reduction in $\rho_c$ is a result of the significantly shortened *c*-axis due to Ca doping, which enhances hopping of electrons (The behavior of $\rho_a$ is qualitatively consistent with that in Ref.29, which, however, does not show c-axis transport and magnetic properties of Ca doped $Sr_2IrO_4$). It is even more surprising that $(Sr_{1-x}Ba_x)_2IrO_4$ with x=0.02 undergoes a sharp insulator-metal transition near 25 K in $\rho_a$ which reduces from 1.03 Ω cm at 25 K to 0.04 Ω cm at 2.0 K (see **Fig.5a inset**). $\rho_c$ still exhibits non-metallic behavior but saturates below 10 K with a drastically reduced magnitude of 11 Ω cm. It needs to be pointed out that high levels of impurity doping (e.g., 15% Ca doping) can cause disorder. However, Ca (or Ba) and Sr are isoelectronic cations, the effect of the disorder on the cation site appears to be less significant than disorder on the on the active Ir site.

The activation energy gap Δ is estimated for a few representative compounds, as shown in **Fig. 5c**. It is clear that Ca and Ba doping significantly narrows the activation energy gap Δ from 107 meV for x = 0 to 13 meV for x = 0.04 or 8 meV for x(Ba) = 0.01. It is noted that the scattering centers can introduce bound states into the gap. They form an impurity band and may change the activation energy, which is not the same quantity as the Mott gap. It is worth mentioning that any impurity (breaking the translational invariance) in an insulator introduces a bound state into the gap; for resonant scattering the levels should be in or close to the gap. Whereas the Sr or Ba states, which are far away from the gap, are immaterial because they are not resonant scattering. In addition, since impurity doping shrinks the lattice parameter V, this may lead to band broadening, which may also help explain the reduced $\Delta$.

While magnetism is a bulk property, a strong increase in the electrical conductivity only requires one connected path through the crystal. Each Ca (Ba) ion substituting a Sr atom locally distorts bonds destroying the coherence of the electronic states. The gap is a manifestation of the coherence of the states at the Fermi level. Hence, the gap is locally reduced due to the impurities favoring the conduction. A small fraction of Ca (Ba) impurities is sufficient to turn the insulator into a bad metal.

### IV. Conclusions

A phase diagram generated based on the data in **Figs.3-5** illustrates the central finding of this work – The drastic changes in the electronic properties require no changes in the Néel temperature in $Sr_2IrO_4$, as shown in **Fig.6a**. This behavior interestingly contrasts that observed in $Sr_2Ir_{1-x}Tb_xO_4$ in which mere 3% $Tb^{4+}$ substitution for Ir effectively suppresses $T_N$ to zero but retains the insulating state, that is, the disappearance of the AFM state accompanies no emergence of a metallic state, as illustrated in **Fig.6b**.[34] Note that $Tb^{4+}$ substitutes for $Ir^{4+}$, the site of the magnetic moment, while Ca and Ba replace Sr ions. A recent theoretical study suggests that the interaction between the magnetic moments on the impurity $Tb^{4+}$ ion and its surrounding $Ir^{4+}$ ions can be described by a "compass" model, i.e., an Ising-like interaction that favors the magnetic moments across each bond to align along the bond direction. This interaction quenches magnetic vortices near the impurities and drives a reentrant transition out of the AFM phase, leading to a complete suppression of the Néel temperature.[35] Nevertheless, the diagrams of **Figs.6a** and **6b** indicate that the magnetic transition plays a nonessential role in the formation of the charge gap in the iridate. For contrast and comparison, we also present a phase diagram for $La^{3+}$ (electron) doped $Sr_2IrO_4$ or $(Sr_{1-x}La_x)_2IrO_4$ ($0 \leq x \leq 0.04$) in **Fig.6c**.[20] Here, $T_N$ decreases with La doping, and completely vanishes at mere $x = 0.04$, where a metallic state is fully established. This behavior is qualitatively consistent with the conventional anticipation although our recent study reveals that there exists an unusual metallic state whose temperature dependence of the resistivity does not follow any known power law down to 50 mK.

It is emphasized that the impurity doping levels in all three cases in **Fig.6** are remarkably low, mostly below 5%, except for 15% for Ca doping. All empirical trends suggest such low doping levels are sufficient to make drastic changes in either the AFM state or the insulating state or both despite the robust AFM and insulating states in the iridate -- Any radical changes in one state (AFM or $J_{eff}=1/2$ state) does not necessarily demand parallel changes in the other ($J_{eff}=1/2$ or AFM state) in the iridate. All results indicate that electron hopping is strongly enhanced as a result of the volume shrinking but the exchange coupling that dictates long-range magnetic order does not seem to be sensitive to the anisotropic compression in the lattice parameters in $(Sr_{1-x}A_x)_2IrO_4$. The vastly different effects of chemical pressure and applied pressure on the AFM state imply the very importance of the in-pane Ir1-O2-Ir1 bond angle in determining the magnetic ground state in $Sr_2IrO_4$.

**Acknowledgments** This work was supported by the US National Science Foundation via grant DMR-1712101.

**Captions**

**Fig. 1**. **(a)** The crystal structure of $Sr_2IrO_4$. Note that the lower panel illustrates the rotation of $IrO_6$ octahedra about the *c*-axis for $Sr_2IrO_4$. **(b)** Representative single crystals of Ca doped $Sr_2IrO_4$.

**Fig.2.** The Ba or Ca concentration x dependence of the lattice parameters **(a)** the *a*-axis, **(b)** the c-axis, **(c)** the unit cell V and **(d)** the Ir1-O2-Ir1 bond angle in the basal plane. Note that the *c*-axis changes more significantly than the *a*-axis with Ca doping.

**Fig.3.** The temperature dependence of **(a)** the *a*-axis magnetization $M_a$ and $\Delta\chi^{-1}$ (right scale), and **(b)** the *c*-axis magnetization $M_c$ at $\mu_oH=1$ T. Note the same scale for $M_a$ and $M_c$ to facilitate comparison. **(c)** The Ba or Ca concentration x dependence of the Néel temperature $T_N$, the Curie-Weiss temperature $\theta_{CW}$ and the effective moment $\mu_{eff}$ (right scale).

**Fig.4.** The isothermal magnetization M(H) for **(a)** the *a*-axis magnetization $M_a$ and **(b)** the *c*-axis magnetization $M_c$ at T=1.8 K for a few representative Ba or Ca doped $Sr_2IrO_4$ compounds. Note the same scale for $M_a$ and $M_c$ to facilitate comparison **(c)** The comparison between $M_a$ (7T) and $M_c$ (7T) at $\mu_oH=7$ T as a function of Ba or Ca concentration x. Note the reduced magnetic anisotropy with Ca doping. **(d)** The schematic illustration of the different effects of chemical pressure and applied pressure [13] on the Néel temperature $T_N$.

**Fig.5.** The temperature dependence of **(a)** the *a*-axis electrical resistivity $\rho_a$, and **(b)** the *c*-axis electrical resistivity $\rho_c$ for a few representative Ba or Ca doped $Sr_2IrO_4$ compounds. The inset in **(a)** is the zoomed-in $\rho$ for Ba doping at x(Ba) = 0.02. Note the sharp insulator-metal transition near

25 K in $\rho_a$. **(c)** The activation energy gap $\Delta$ estimated from the data in **(a)** for a few representative compounds. Note the rapid changes in $\Delta$ with Ba or Ca doping.

**Fig.6.** The phase diagram for **(a)** $(Sr_{1-x}A_x)_2IrO_4$ (where A=Ba or Ca), **(b)** $Sr_2Ir_{1-x}Tb_xO_4$, and **(c)** for $(Sr_{1-x}La_x)_2IrO_4$. Note (1) M = metal, and I = insulator, AFM-I = AFM insulator, PM-I = paramagnetic insulator, and PM-M = paramagnetic metal; (2) Color code: The white color and dark blue represent an insulating state and a metallic state, respectively; color between white and dark blue represents states that is between an insulating state and a metallic state; the darker the blue is the more metallic the system becomes.

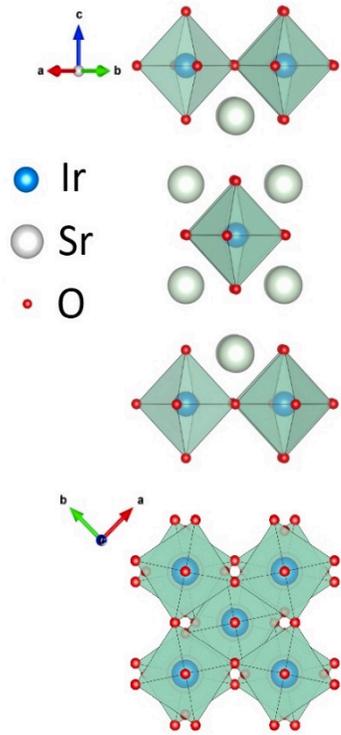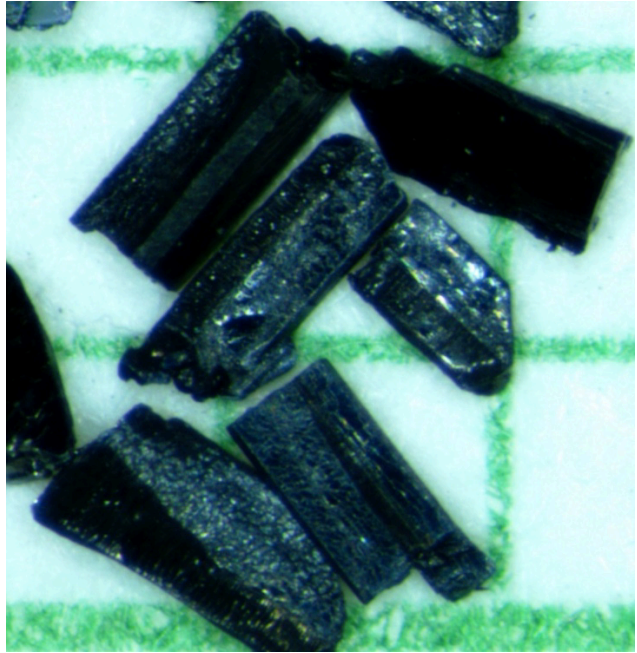

Fig.1

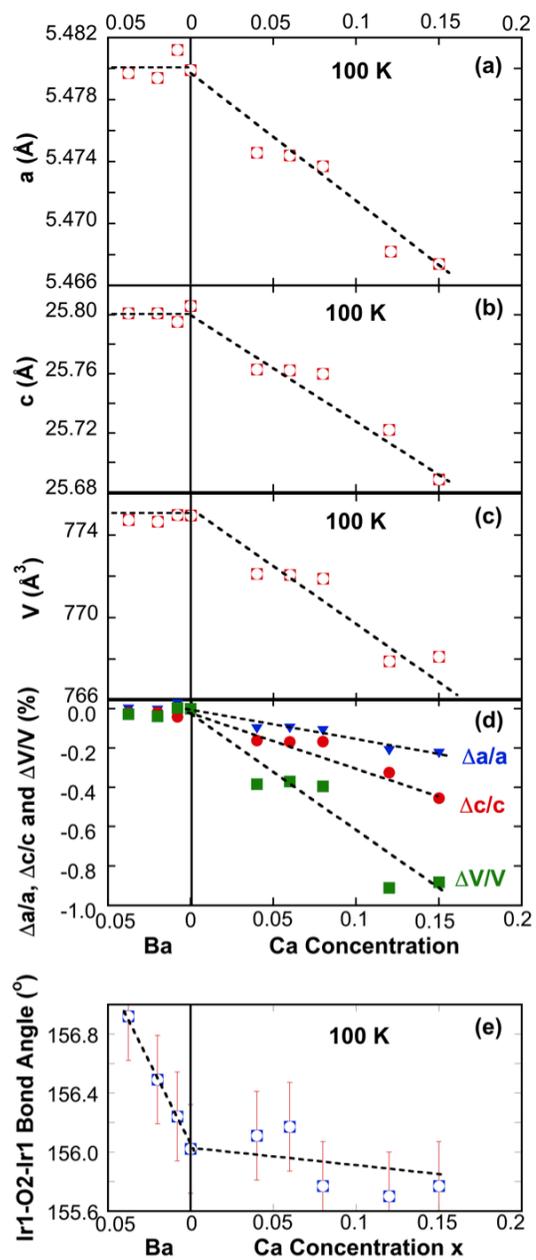

Fig. 2

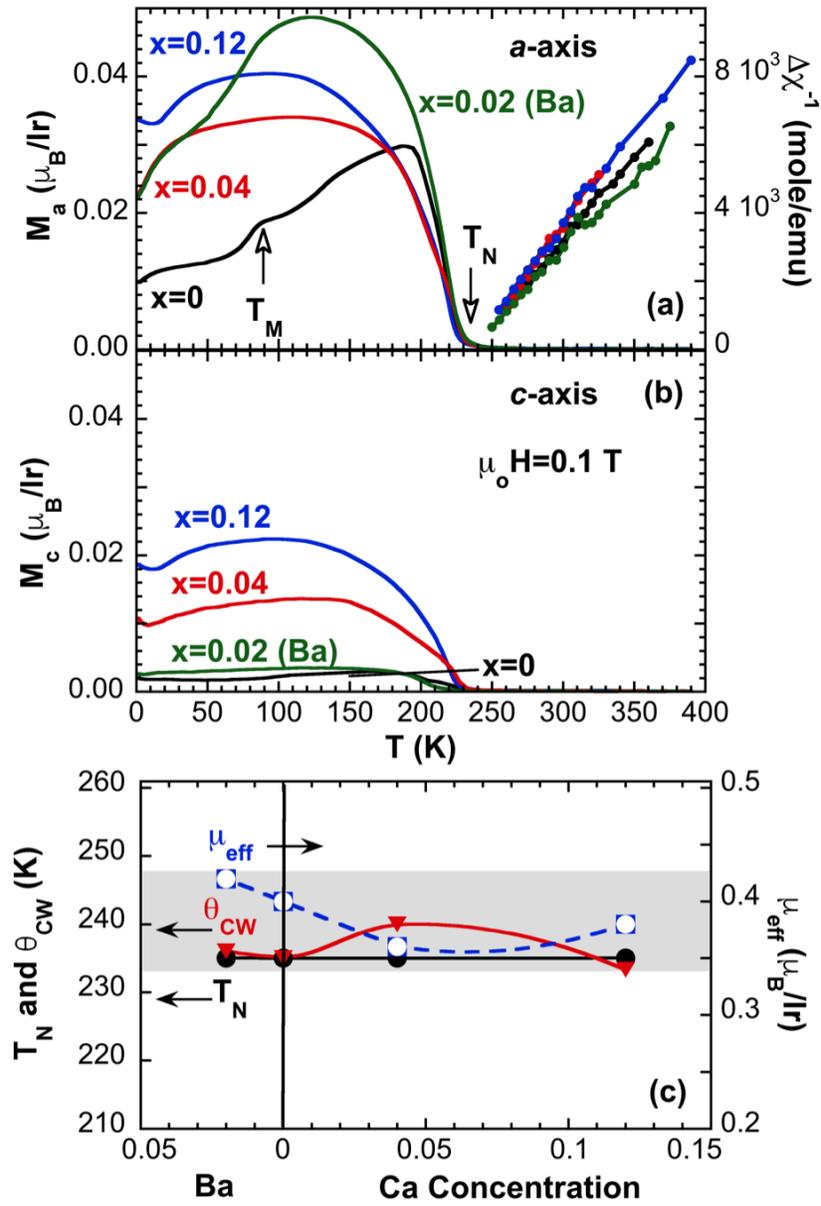

Fig. 3

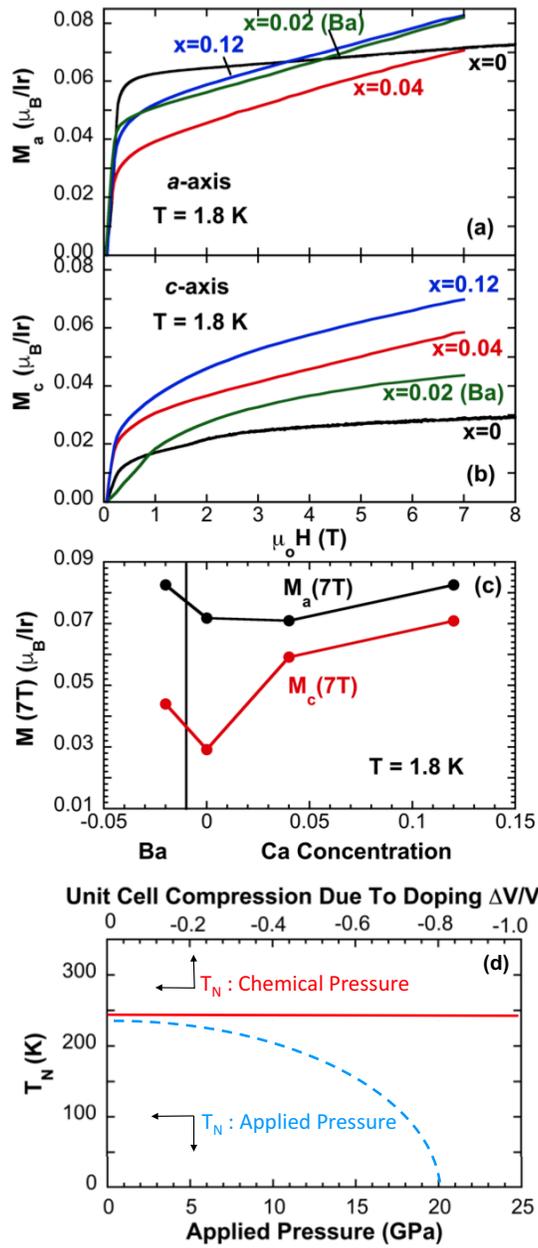

Fig.4

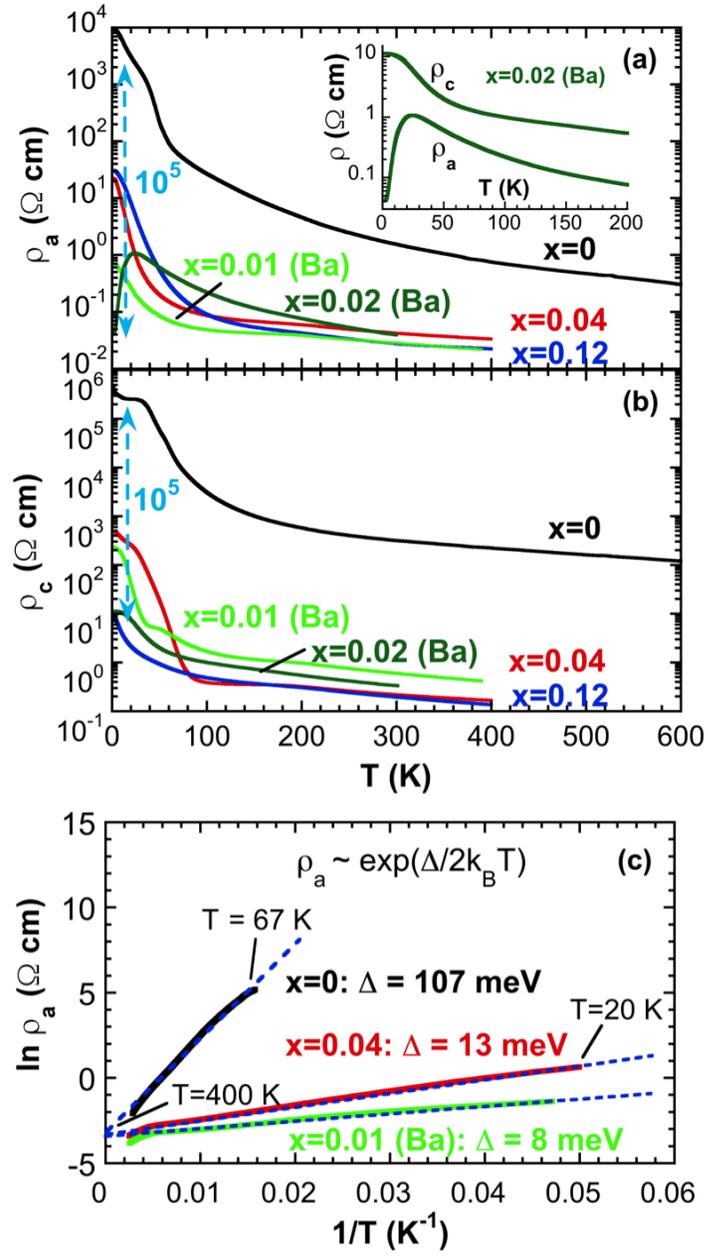

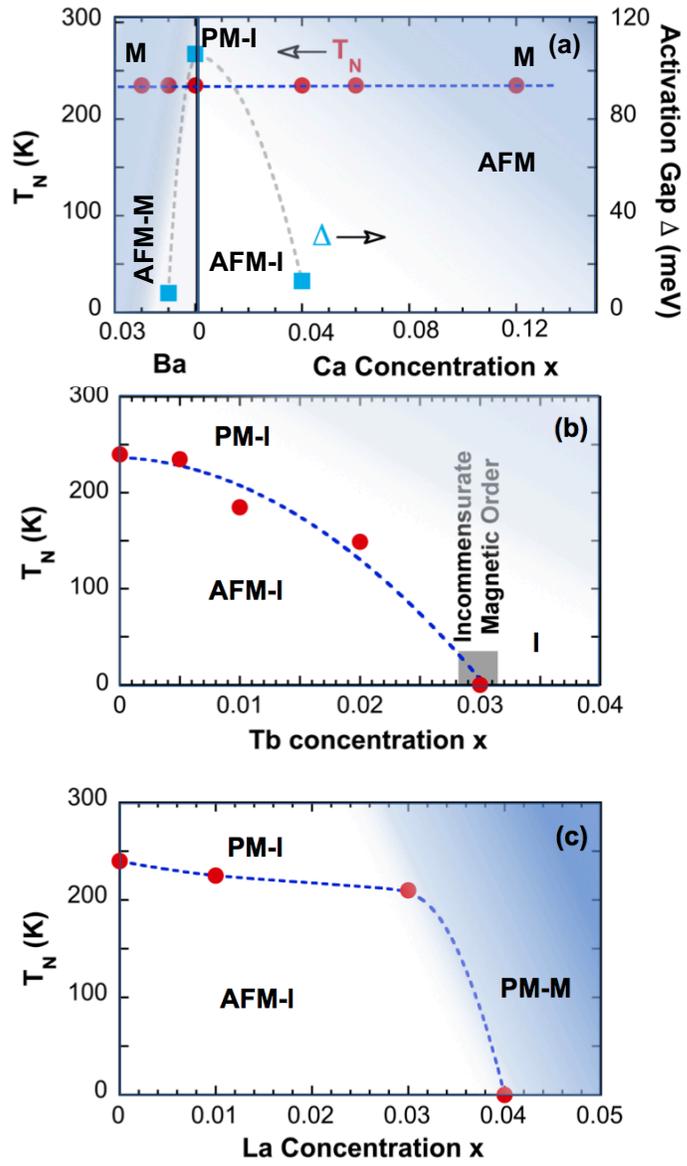

Fig. 6.